\documentclass[aps,epsf,twocolumn,showpacs,superscriptaddress]{revtex4}
\usepackage{amsmath}
\usepackage{epsfig}
\usepackage{multirow}

\begin{document}

\title{Universality in the three-dimensional random-field Ising model}

\author{Nikolaos G. Fytas}

\affiliation{Departamento de F\'{i}sica Te\'{o}rica I,
Universidad Complutense, E-28040 Madrid, Spain.}

\affiliation{Applied Mathematics Research Centre, Coventry
University, Coventry, CV1 5FB, United Kingdom.}

\author{V\'{i}ctor Mart\'{i}n-Mayor}

\affiliation{Departamento de F\'{i}sica Te\'{o}rica I,
Universidad Complutense, E-28040 Madrid, Spain.}

\affiliation{Instituto de Biocomputaci\'on and
  F\'{\i}sica de Sistemas Complejos (BIFI), E-50009 Zaragoza, Spain.}

\begin{abstract}
We solve a long-standing puzzle in Statistical Mechanics of
disordered systems. By performing a high-statistics simulation of
the $D=3$ random-field Ising model at zero temperature for
different shapes of the random-field distribution, we show that
the model is ruled by a single universality class. We compute the
complete set of critical exponents for this class, including the
correction-to-scaling exponent, and we show, to high numerical
accuracy, that scaling is described by two independent exponents.
Discrepancies with previous works are explained in terms of strong
scaling corrections.
\end{abstract}

\pacs{75.10.Nr, 02.60.Pn, 75.50.Lk}

\maketitle

The random-field Ising model (RFIM), is one of the simplest and most
investigated models for collective behavior in the presence of quenched
disorder~\cite{natterman98}. In spite of its simplicity, many problems in
Condensed Matter Physics can be studied through the RFIM: diluted
antiferromagnets in a field~\cite{belanger98}, colloid-polymer
mixtures~\cite{vink06,annunziata12}, colossal magnetoresistance
oxides~\cite{dagotto05,burgy01} (more generally, phase-coexistence in the
presence of quenched disorder~\cite{cardy97,fernandez08,fernandez12}),
non-equilibrium phenomena such as the Barkhausen noise in magnetic
hysteresis~\cite{sethna93,perkovic99} or the design of switchable magnetic
domains~\cite{silevitch10}, etc.

On the theoretical side, a scaling picture is
available~\cite{imry75,villain84,bray85,fisher86}. The
paramagnetic-ferromagnetic phase transition is ruled by a fixed-point [in the
  Renormalization-Group (RG) sense] at temperature
$T=0$~\cite{natterman98}. The spatial dimension $D$ is replaced by $D-\theta$,
in hyperscaling relations ($\theta\approx D/2$). Nevertheless, many expect
only two independent exponents~\cite{aharony76,gofman93,natterman98}, as in
standard phase transitions (see e.g.~\cite{amit05}).

Unfortunately, establishing the scaling picture is far from
trivial. Perturbation theory predicts that, in $D=3$, the ferromagnetic phase
disappears upon applying the tiniest random field~\cite{young77}. Even if the
statement holds at all orders in perturbation theory~\cite{parisi79}, the
ferromagnetic phase \emph{is} stable in
$D=3$~\cite{bricmont87}. Nonperturbative phenomena are obviously at
play~\cite{parisi94,parisi02}. Indeed, it has been suggested that the scaling
picture breaks down because of spontaneous supersymmetry breaking, implying
that more than two critical exponents are needed to describe the phase
transition~\cite{tissier11}.

On the experimental side, a particularly well researched realization of the
RFIM is the diluted antiferromagnet in an applied magnetic
field~\cite{belanger98}. Yet, there are inconsistencies in the determination
of critical exponents. In neutron scattering, different parameterizations of
the scattering line-shape yield mutually incompatible estimates of the thermal
critical exponent, namely $\nu=0.87(7)$~\cite{slanic99} and
$\nu=1.20(5)$~\cite{ye04}. Moreover, the anomalous dimension
$\eta=0.16(6)$~\cite{slanic99} violates hyperscaling bounds (at least if one
believes experimental claims of a divergent specific
heat~\cite{belanger83,belanger98b}). Clearly, a reliable parametrization of
the line-shape would be welcome. This program has been carried out for
simpler, better understood problems~\cite{martinmayor02}. Unfortunately, we do
not have such a strong command over the RFIM universality class.

The RFIM has been also investigated by means of numerical
simulations. However, typical Monte Carlo schemes get trapped into
local minima with escape time exponential in $\xi^\theta$ ($\xi$
is the correlation length). Although sophisticated improvements
have appeared~\cite{fernandez11,fytas08,ahrens}, these simulations
produced low-accuracy data because they were limited to linear
sizes $L\leq 32$.  Larger sizes can be achieved at $T=0$, through
the well-known mapping of the ground state to the maximum-flow
optimization
problem~\cite{ogielski86,auriac,sourlas,auriac97,swift,hartmann99,hartmann01,middleton02a,middleton02b,machta03,machta05}.
Yet, $T=0$ simulations lack many tools, standard at $T>0$. In
fact, the numerical data at $T=0$ and their finite-size scaling
analysis mostly resulted in strong universality
violations~\cite{sourlas,auriac97,swift,hartmann99,ahrens}.

Here we show that the $D=3$ RFIM is ruled by a single universality
class, by considering explicitly four different models that belong
to it. To this end, we perform high-statistics $T=0$ simulations
of the model and we introduce a fluctuation-dissipation formalism
in order to compute connected and disconnected correlation
functions. Another asset of our implementation is the use of
phenomenological renormalization~\cite{night,bal96}, that allows
us to extract effective size-dependent critical exponents, whose
size evolution can be closely followed. Although the four models
differ in their prediction for finite sizes, we show that, after a
proper consideration of the scaling corrections, we can
extrapolate to infinite-limit size, finding consistent results for
all of them.

\begin{figure}%[htbp]
\includegraphics*[width=8 cm]{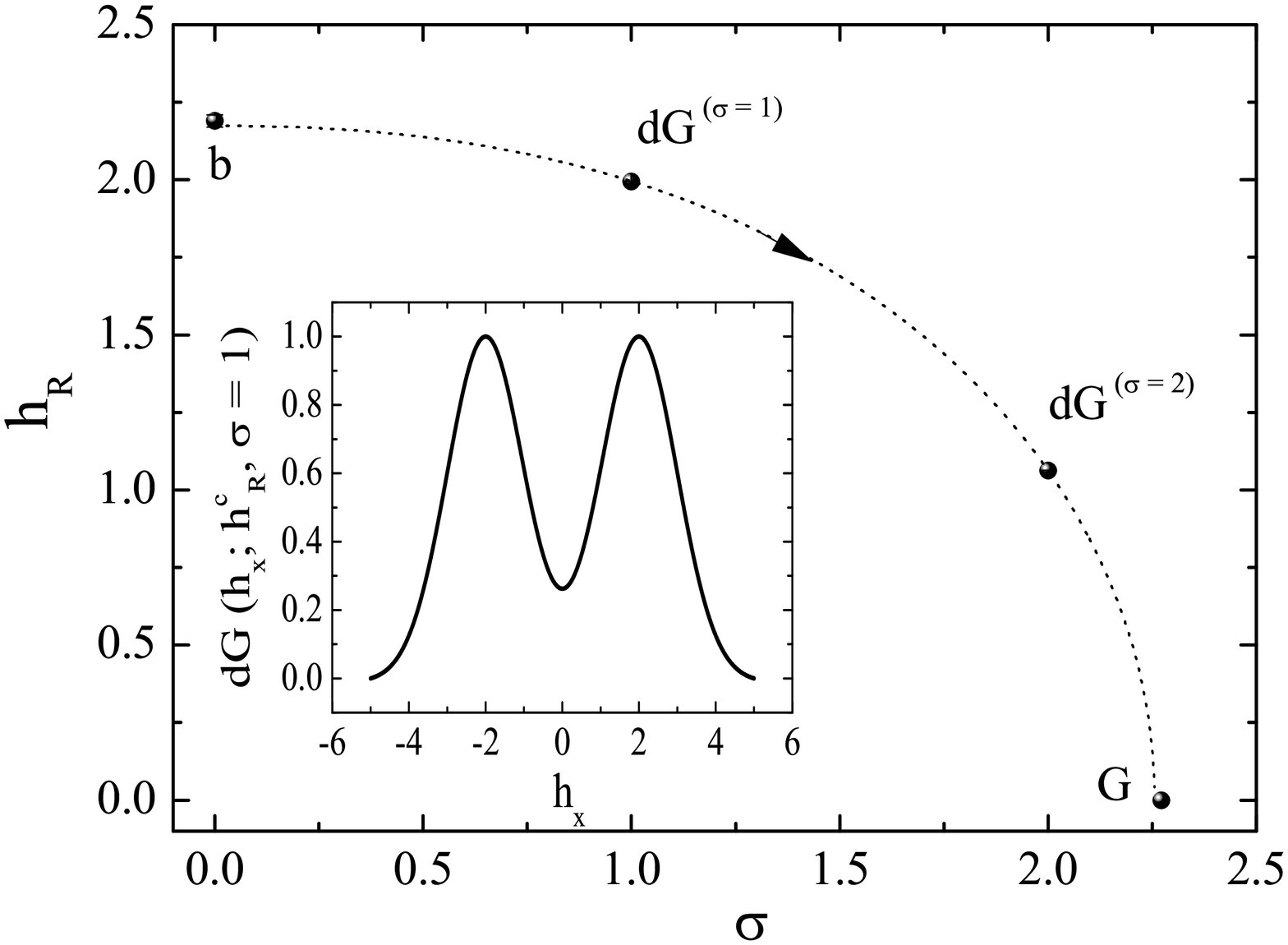}
\caption{\label{fig:phase-diagram} Phase diagram of the double Gaussian RFIM,
  Eq.~\eqref{eq:dGaussian}, at $T=0$. The dotted critical line (a simple guide
  to the eye) separates the paramagnetic phase (large $\sigma$, $h_{R}$) from
  the ordered phase (low $\sigma$, $h_{R}$). Transition points are computed
  here (data from Table~\ref{tab:FSS_full} in the Appendix), but for the limit $\sigma=0$ which
  corresponds to binary random fields~\cite{hartmann99}. The arrow along the
  critical line indicates the RG flow. {\bf Inset}: bimodal shape of the
  critical double Gaussian distribution~(\ref{eq:dGaussian}) with $\sigma=1$.}
\end{figure}

Our $S_{x}=\pm 1$ spins are on a cubic lattice with size $L$ and
periodic boundary conditions. The Hamiltonian is
\begin{equation}
\label{eq:H} \mathcal{H}=-J\sum_{\langle
x,y\rangle}S_{x}S_{y}-\sum_{x}h_{x}S_{x}.
\end{equation}
$J=1$ is the nearest-neighbors' ferromagnetic interaction. Independent
quenched random fields $h_{x}$ are extracted from one of the following double
Gaussian (dG) or Poissonian (P) distributions (with parameters $h_R$,
$\sigma$):
\begin{eqnarray}
{\rm
  dG}^{(\sigma)}(h_{x};h_{R})&=&\frac{1}{\sqrt{8\pi\sigma^2}}\big[e^{-\frac{(h_{x}-h_{R})^{2}}{2\sigma^{2}}}+
  e^{-\frac{(h_{x}+h_{R})^{2}}{2\sigma^{2}}}\big],\ \ \label{eq:dGaussian}\\
{\rm P}(h_{x};\sigma)&=&\frac{1}{2|\sigma|}
e^{-|h_{x}| / \sigma}\,.\label{eq:Poisson}
\end{eqnarray}
The limiting cases $\sigma=0$ and $h_{R}=0$ of Eq.~(\ref{eq:dGaussian})
correspond to the well-known bimodal (b) and Gaussian (G) distributions,
respectively. In the P and G cases the strength of the random fields is
parameterized by $\sigma$, while in the dG case we shall take
$\sigma=1$ and $2$, and vary $h_R$. The phase diagram for the double Gaussian
distribution is sketched in Fig.~\ref{fig:phase-diagram}. Note the bimodal
shape of Eq.~\eqref{eq:dGaussian} for $\sigma=1$, with peaks near $\pm h_R$.

An instance of the random fields $\{h_x\}$ is named a sample.  The only
relevant spin configurations at $T=0$ are ground states, which are
non-degenerate for continuous random-field
distributions~\cite{bastea}. Thermal mean values are denoted as $\langle\cdots
\rangle$. The subsequent average over samples is indicated by an
over-line (e.g., for the magnetization density $m=\sum_x s_x/L^D$ we consider
both $\langle m\rangle$ and $\overline{\langle m \rangle}$).

We considered four disorder distributions: P, G and dG with $\sigma=1,2$. We
obtained the ground-states using the push-relabel algorithm~\cite{tarjan}. We
implemented in C the algorithm in~\cite{middleton02a,middleton02b}, with
periodic global updates.  Our lattices sizes were $L=12,16,24,32,48,64,96,128$
and $192$ [$16\leq L \leq 128$ for dG$^{(\sigma=1)}$ and $12\leq L \leq 128$
  for dG$^{(\sigma=2)}$].  For each $L$, we averaged over $10^7$ samples
[$5\times 10^7$ samples for dG$^{(\sigma=1)}$]. Previous studies were limited
to $\sim 10^{4}$ samples~\cite{hartmann01,middleton02a}.

We have generalized the fluctuation-dissipation formalism of~\cite{soffer} to
compute connected $G_{xy}=\overline{\partial [\langle S_{x}\rangle]/\partial
  h_{y}}$ and disconnected $G_{xy}^{\rm (dis)}=\overline{\langle
  S_{x}S_{y}\rangle}$ correlation functions. We compute from them the
second-moment connected ($\xi$) and disconnected ($\xi^{\rm (dis)}$)
correlation lengths~\cite{amit05,cooper}.

\begin{figure}%[htbp]
\includegraphics*[width=8 cm]{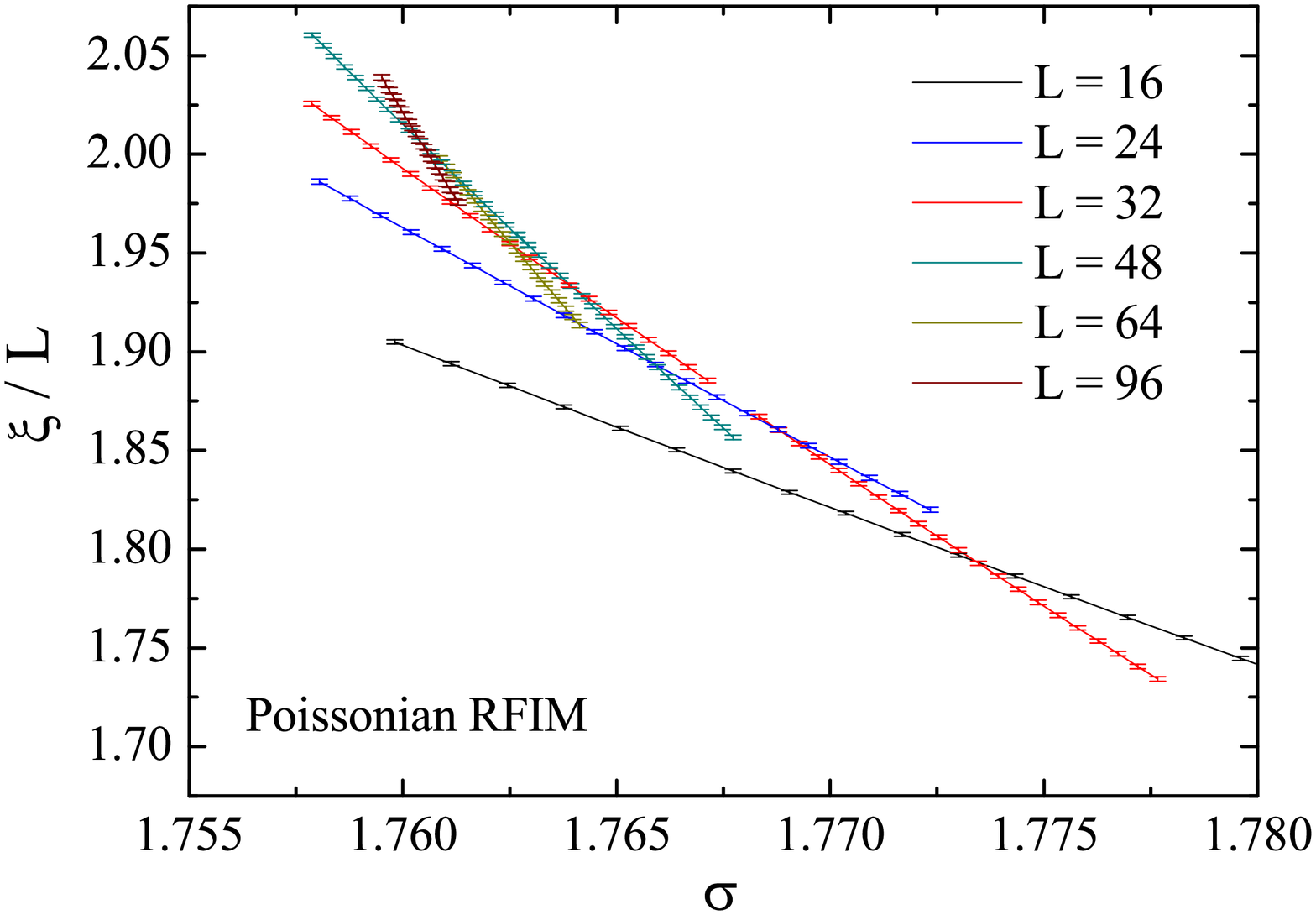}
\caption{\label{fig:crossing} (Color online) For several system sizes, we show
  $\xi/L$ vs. the strength of the Poissonian random field $\sigma$ [see
    Eq.~\eqref{eq:Poisson}] ($\xi$: correlation length from the connected
  correlator, $L$: system size). Lines join data obtained from reweighting
  extrapolation, Eq.~\eqref{eq:reweighting} (discontinuous lines of the same
  color come from independent simulations).  In the large-$L$ limit, $\xi/L$
  is $L$-independent at the critical point $\sigma^{c}$. In the quotients
  method, we consider the $\xi/L$ curves for pair of lattices $(L,2L)$ and
  seek the $\sigma$ where they \emph{cross}. This crossing is employed for
  computing effective, $L$-dependent critical exponents with
  Eq.~\eqref{eq:QO}.}
\end{figure}

We have also extended reweighting methods from percolation
studies~\cite{harris94,bal98}. From a single simulation, we extrapolate the
mean value of observables to nearby parameters of the disorder distribution
(we varied $\sigma$ for the G and P distributions, see
Fig.~\ref{fig:crossing}, and $h_{R}$ for the dG case). Computing derivatives
with respect to $\sigma$ or $h_R$ is straightforward. Consider, for instance,
the P case (see~\cite{prepare} for other distributions). Let ${\cal D}= \sum_x
(|h_x|-\sigma)/\sigma^2$. The connected correlation function is
$G_{xy}=\overline{h_y\langle S_x\rangle/(|h_y|\sigma)})$, while the
$\sigma$-derivative and the reweighting-extrapolation to $\sigma+\delta\sigma$
of a generic observable $O$ are
\begin{eqnarray}\label{eq:reweighting}
D_\sigma \overline{\langle O \rangle}_\sigma=  \overline{\langle O {\cal D} \rangle}_\sigma&,& \overline{\langle O\rangle}_{\sigma+\delta\sigma}=
\overline{\langle O {\cal R}\rangle\, }_\sigma\quad \text{with}
\\\nonumber
{\cal  R}= \mathrm{exp}\bigg[{\cal D}
  \frac{\sigma\delta\sigma}{\sigma+\delta\sigma}&+& L^D \bigg(\mathrm{log}
  \frac{\sigma}{\sigma+\delta\sigma} + \frac{\sigma+\delta\sigma}
       {\sigma}\bigg)\bigg]\,.
\end{eqnarray}

To extract the value of critical points, critical exponents, and dimensionless
quantities, we employ the quotients method~\cite{night,bal96,amit05}. We
compare observables computed in pair of lattices $(L,2L)$. We start imposing
scale-invariance by seeking the $L$-dependent critical point: the value of
$\sigma$ ($h_R$ for the dG), such that $\xi_{2L}/\xi_L=2$ (i.e. the
\emph{crossing} point for $\xi_L/L$, see Fig~\ref{fig:crossing}). Now, for
dimensionful quantities $O$, scaling in the thermodynamic limit as
$\xi^{x_O/\nu}$, we consider the quotient $Q_O=O_{2L}/O_L$ at the
crossing. For dimensionless magnitudes $g$, we focus on $g_{2L}$.  In either
case, one has:
\begin{equation}\label{eq:QO}
Q_O^\mathrm{cross}=2^{x_O/\nu}+\mathcal{O}(L^{-\omega})\,,\
g_{(2L)}^\mathrm{cross}=g^\ast+\mathcal{O}(L^{-\omega})\,,
\end{equation}
where $x_O/\nu$, $g^\ast$ and the scaling-corrections exponent
$\omega$ are universal.  Examples of dimensionless quantities are
$\xi/L$, $\xi^\mathrm{(dis)}/L$ and $U_4=\overline{\langle
m^4\rangle}/ \overline{\langle
  m^2\rangle}^2$. Instances of dimensionful quantities are the derivatives of
$\xi$, $\xi^{\rm (dis)}$ ($x_\xi=1+\nu$), the connected and
disconnected susceptibilities $\chi$ and $\chi^{\rm (dis)}$
[$x_\chi= \nu(2-\eta)$,
  $x_{\chi^{\rm (dis)}}= \nu(4-\bar\eta)$], and the ratio
$U_{22}=\chi^\mathrm{(dis)}/\chi^2$
[$x_{U_{22}}=\nu(2\eta-\bar\eta)$].

\begin{table}
\caption{\label{tab:FSS} For our four field distributions, size-dependent
  critical exponents of the $D=3$ RFIM  as computed from the quotients method.}
\begin{tabular}{llccc}
\hline \hline
Distr. & $(L_{1},L_{2})$ & $\nu$ & $\eta$ & $2 \eta-\bar{\eta}$ \\
\hline
G  &$(16,32)$   &  1.48(3)  & 0.5168(6) & 0.0038(11)  \\
   &$(24,48)$   &  1.45(3)  & 0.5155(5) & 0.0022(11) \\
   &$(32,64)$   &  1.36(4)  & 0.5150(5) & 0.0019(10) \\
   &$(48,96)$   &  1.43(6)  & 0.5154(5) & 0.0033(9) \\
   &$(64,128)$  &  1.38(9)  & 0.5142(5) & 0.0014(10) \\
   &$(96,192)$  &  1.38(11) & 0.5144(5) & 0.0021(11) \\
\hline \hline
dG$^{(\sigma=1)}$&$(16,32)$   &   3.04(14)  & 0.5035(7) & 0.0016(15) \\
                 &$(24,48)$   &   2.26(9)   & 0.5083(7) & 0.0034(14) \\
                 &$(32,64)$   &   1.87(8)   & 0.5093(7) & 0.0010(13) \\
                 &$(48,96)$   &   1.56(9)   & 0.5121(7) & 0.0026(14) \\
                 &$(64,128)$  &   1.67(12)  & 0.5125(8) & 0.0015(17)  \\
\hline \hline
dG$^{(\sigma=2)}$&$(16,32)$   & 1.48(5)   &  0.5154(6) &  0.0020(12) \\
                 &$(24,48)$   & 1.50(6)   &  0.5151(7) &  0.0020(13) \\
                 &$(32,64)$   & 1.41(8)   &  0.5142(7) &  0.0004(13) \\
                 &$(48,96)$   & 1.36(10)  &  0.5148(7) &  0.0024(14) \\
                 &$(64,128)$  & 1.31(11)  &  0.5154(6) &  0.0041(13)  \\
\hline \hline
P  &$(16,32)$   &   1.20(2)    & 0.5183(9) & -0.0006(19) \\
   &$(24,48)$   &   1.26(3)    & 0.5168(8) & 0.0011(17) \\
   &$(32,64)$   &   1.30(4)    & 0.5153(8) & 0.0005(17) \\
   &$(48,96)$   &   1.37(7)    & 0.5143(9) & 0.0004(18) \\
   &$(64,128)$  &   1.33(7)    & 0.5148(8) & 0.0024(16) \\
   &$(96,192)$  &   1.43(13)   & 0.5146(8) & 0.0026(17) \\
   \hline \hline
\end{tabular}
\end{table}

\begin{figure}%[htbp]
\includegraphics*[width=8 cm]{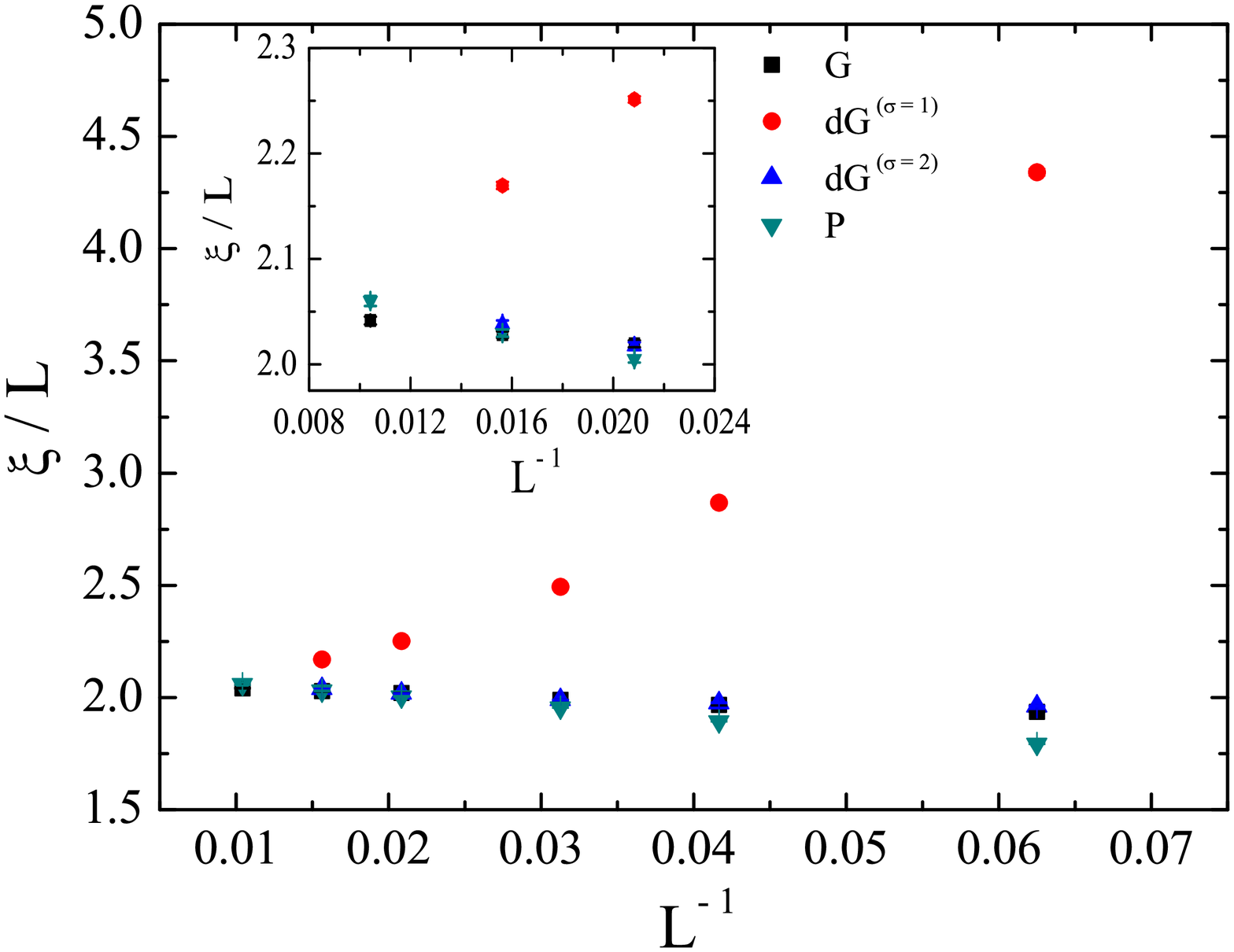}
\caption{\label{fig:inspection} (Color online) Inspection of the
  size-dependence of universal quantities. We show the universal ratio $\xi/L$
  versus $1/L$, as computed at the corresponding crossing points (see
  Fig~\ref{fig:crossing}) for the four disorder distributions considered in
  this work. The inset is an enlargement for $L\geq 48$.}
\end{figure}

The application of Eq.~\eqref{eq:QO} to our four random-field
distributions is summarized in Table~\ref{tab:FSS} and
Figs.~\ref{fig:inspection} and~\ref{fig:omega} (the numerical
values are listed in Table~\ref{tab:FSS_full} in the Appendix). We
start inspecting $\xi/L$ in Fig.~\ref{fig:inspection}. At fixed
$L$, the dependence on the distribution is substantial. However,
the strong size evolution suggests a common $L\to\infty$ limit.
The behavior of the critical exponents, $\xi^\mathrm{(dis)}/L$ and
$U_4$ is similar.

\begin{table}
\caption{\label{tab:quotients} Extrapolations to $L=\infty$ for the critical
  $\xi/L$, $\xi^{\mathrm{(dis)}}/L$, $U_4$ and the exponents of the $D=3$
  RFIM, (finite-$L$ data from Table~\ref{tab:FSS_full} in the Appendix). We
  perform joint fits for sizes $L\geq L_{\rm min}$ to polynomials in
  $L^{-\omega}$, imposing common extrapolations for all four random-field
  distributions. The $\chi^2$ figure of merit was computed with the full
  covariance matrix (${\rm dof}$: number of degrees of freedom in the
  fit). For the exponent $\nu$, we considered derivatives of both $\xi$ and
  $\xi^{\mathrm{(dis)}}$. The error induced by the uncertainty in $\omega$ is
  given as a second error estimate. The extrapolation of the critical points
  slightly differs~\cite{CRITICAL}.}
\begin{ruledtabular}
\begin{tabular}{r@{\,=\,}lccc}
\multicolumn{2}{c}{Extrapolation} & $\chi^{2}/{\rm dof}$ & $L_{\rm min}$ & order in $L^{-\omega}$\\
\hline
$(\xi/L)|_{L=\infty}$&$2.08(13)$               & \multirow{4}{*}{$18.8/14$}     & \multirow{4}{*}{24}  & \multirow{4}{*}{third}     \\
$(\xi^{\rm (dis)}/L)|_{L=\infty}$&$8.4(8)$        &     &     &       \\
$U_{4}|_{L=\infty}$&$1.0011(18)$                  &     &     &       \\
$\omega$&$0.52(11)$                   &     &     &       \\
\hline $\nu|_{L=\infty}$&$1.38(10)(0.03)$                      &
$12.5/10$    &   32  &
first     \\
$\eta|_{L=\infty}$&$0.5153(9)(2)$                    & $10.0/9$        &   32  &
first     \\
$(2\eta-\bar{\eta})|_{L=\infty}$&$0$ (fixed)      & $18.3/18$    &
16  &
first     \\
$(2\eta-\bar{\eta})|_{L=\infty}$&$0.0026(9)(1)$      & $10.5/17$    &   16  &
first     \\
\hline
$\sigma^{c}[{\rm G}]$&$2.27205(18)(4)$               & $3.1/3$      &   16  &
second    \\
$h_{R}^{c}[{\rm dG}^{(\sigma=1)}]$&$1.9955(6)(24)$  & $2.5/1$      &   24  &
second    \\
$h_{R}^{c}[{\rm dG}^{(\sigma=2)}]$&$1.0631(7)(10)$  & $0.7/2$ & 16
&
second    \\
$\sigma^{c}[{\rm P}]$&$1.7583(2)(2)$               & $3.0/3$      &   16  &  second    \\
\end{tabular}
\end{ruledtabular}
\end{table}

\begin{figure}%[htbp]
\includegraphics*[width=7 cm]{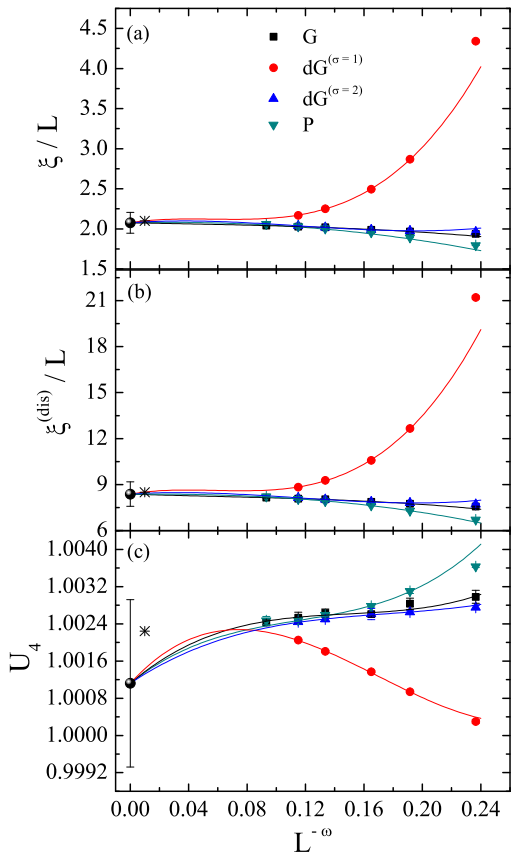}
\caption{\label{fig:omega} (Color online) Computation of the
  corrections-to-scaling exponent $\omega$, see Eq.~\eqref{eq:QO}, by means of
  a joint fit for $\xi/L$ {\bf (a)}, $\xi^{\rm (dis)}/L$ {\bf (b)}, and
  $U_{4}=\overline{\langle m^4\rangle}/ \overline{\langle m^2\rangle}^2$ {\bf
    (c)}, see Table~\ref{tab:quotients} and~\cite{JOINT}. Black circles
  depicted at $L^{-\omega}=0$ are our extrapolations to $L=\infty$. Stars
  denote extrapolations obtained using only the diagonal terms of the
  covariance matrix.}
\end{figure}

In order to extrapolate to $L\to\infty$, one fits the data of
Table~\ref{tab:FSS} to polynomials in $L^{-\omega}$. Although the
procedure is standard~\cite{bal98}, it has not been attempted
before for the RFIM. Our extrapolations are documented in
Fig.~\ref{fig:omega} and Table~\ref{tab:quotients}. Several
comments are in order:
\begin{itemize}

\item For dimensionless quantities we needed a third-order polynomial in
  $L^{-\omega}$ (only a subclass of the sub-leading
  corrections-to-scaling~\cite{amit05}). Leading-order corrections sufficed
  for critical exponents.

\item We are not aware of any other computation of $\omega$,
$\xi/L$, $\xi^\mathrm{(dis)}/L$, and $U_4$. All of them are
universal.

\item The full critical line belongs to a single universality
class (justifying \emph{a posteriori} the RG flow in
Fig.~\ref{fig:phase-diagram}). The fixed point at $\sigma=0$
(bimodal RFIM)~\cite{sourlas,auriac97,swift,hartmann99} has no
basin of attraction.

\item The two-exponent scaling scenario~\cite{natterman98}
  ($2\eta=\bar{\eta}$) is supported by our data. Violations are at most $\sim
  0.002$, much smaller than predicted in~\cite{tissier11}.

\item Our $\nu$ estimation is similar to modern
  computations~\cite{hartmann01,middleton02b,machta03,machta05}.  For the
  anomalous dimension, we note $\eta=0.50(3)$~\cite{hartmann01}.  With some
  caution, we quote as well~\cite{yllanes11}.  Our errors for $\nu$ are larger
  than for $\eta$ or $\bar\eta$ because we compute derivatives as connected
  correlations, Eq.~\eqref{eq:reweighting}.

\item The two-exponent scenario implies for the hyperscaling violation
  exponent $\theta=2-\eta=1.4847(11)$ (recall
  $\theta=1.469(20)$~\cite{fernandez11} or $1.49(3)$~\cite{middleton02a}).

\item For the Gaussian RFIM, our critical point compares favorably
to the value $2.270(4)$ of Ref.~\cite{middleton02a}.
\end{itemize}

Where do we stand, at this point? Clearly, our confirmation of
Universality provides motivation to undertake an ambitious
field-theoretic study of the RFIM. The task is difficult, though.
The direct approach of Refs.~\cite{young77,parisi79} fails because
the supersymmmetry is spontaneously broken as soon as
$\theta\!=2$. The phenomenological two-exponent scaling
picture~\cite{aharony76,gofman93} seems highly successful, but it
has not yet being derived from field theory. However, it does
suggest a systematic way of computing critical exponents: study
the ferromagnetic Ising model in the absence of disorder, but with
a non-integer spatial dimension $d'=D-\theta$~\cite{gofman93}
($d'\approx 1.5153(11)$ according to our results for $D=3$). The
computation has been attempted within the $\epsilon$-expansion,
where $\epsilon =4-d'$. The results up to order $\epsilon^5$ are
encouraging, but not very accurate~\cite{jolicoeur97}. On the
other hand, a first-principles approach, the functional
Renormalization Group (fRG)~\cite{wetterich93}, does explain the
spontaneous breaking of the supersymmetry~\cite{tissier11}. Yet,
the fRG prediction for the critical exponents is rather crude.
Furthermore, the smallness of $2\eta-\bar\eta$ (see
Table~\ref{tab:quotients}) remains unexplained. Fortunately, fRG
computations can be systematically improved through the
parametrization of the effective action. In this way, high
accuracy has been reached for non-disordered
systems~\cite{benitez12}. Refinements are probably feasible as
well for the RFIM~\cite{tarjus13}.

In conclusion, we have shown that the universality class of the
RFIM is independent of the form of the implemented random-field
distribution, in disagreement with the current opinion in the
literature~\cite{fytas08,sourlas,auriac97,swift,hartmann99,ahrens}
and with the early predictions of mean-field
theory~\cite{aharony78}.  To reach this conclusion, we had to
identify and control the role of scaling corrections, the Achilles
heel in the study of the RFIM (this problem was emphasized in the
pioneering work of~\cite{ogielski86}, but it was overlooked in
subsequent investigations). On technical parts, we have developed
a fluctuation-dissipation formalism that allowed us to compute
correlation functions and to apply phenomenological
renormalization. We have also adapted the approach
of~\cite{janke97} to study the scaling of the energy (see
Table~\ref{tab:specheat} in the Appendix), checking that our data
are compatible with modified hyperscaling~\cite{prepare} (a rather
slippery problem~\cite{hartmann01}). Hence, several contradictions
of previous works have been resolved in a consistent picture,
paving the way to more sophisticated, experimentally relevant
computations.

We were partly supported by MICINN, Spain, through research contracts
No. FIS2009-12648-C03, FIS2012-35719-C02-01. Significant allocations of
computing time were obtained in the clusters \emph{Terminus} and
\emph{Memento} (BIFI).  We are grateful to D. Yllanes and, especially, to
L.A. Fern\'{a}ndez for substantial help during several parts of this work.  We
also thank A. Pelissetto and G. Tarjus for useful correspondence.

\newpage

\appendix

\section{The critical exponent of the specific heat}

In order to compute the specific-heat's critical exponent
$\alpha$, we consider the exchange part of the Hamiltonian (see
main text for details):
\begin{equation}
E_\mathrm{exch} = -J\sum_{\langle x,y\rangle}S_{x}S_{y}\,.
\end{equation}
Finite-size scaling tell us that, at the critical point,
\begin{equation}\label{eq:scaling-C}
\overline{\langle E_\mathrm{exch}\rangle} \sim A + B
L^{\frac{\alpha-1}{\nu}}\,
\end{equation}
where $A$ and $B$ are non-universal constants. Since $\alpha-1$ is
negative, Eq.~\eqref{eq:scaling-C} is dominated by the
non-divergent back ground $A$, forcing us to modify the standard
phenomenological renormalization. We get rid of $A$ by considering
\emph{three} lattices. At variance with the standard two-lattices
phenomenological renormalization, statistical errors are
significantly amplified by the reweighting extrapolation. Hence,
we have preferred to carry out an independent set of simulations
for parameters corresponding to the crossing points identified in
the main manuscript and listed on Table~\ref{tab:FSS_full}. Our
results for $(\alpha-1)/\nu$ are given in
Table~\ref{tab:specheat}. Details will be provided elsewhere (see
N.G. Fytas and V. Mart\'{i}n-Mayor, manuscript in preparation).

\begin{table}[h]
\begin{ruledtabular}
\caption{\label{tab:specheat} Effective critical exponent ratio
  $(\alpha-1)/\nu$ using a three-lattice size variant
  $(L_{1},L_{2},L_{3})=(L,2L,4L)$ of the quotients method.}
\begin{tabular}{llc}
Distr. & $(L_{1},L_{2},L_{3})$ & $(\alpha-1)/\nu$\\
\hline
G    &$(12,24,48)$    &   -0.758(11)          \\
     &$(16,32,64)$    &   -0.793(17)         \\
     &$(24,48,96)$    &   -0.860(30)         \\
     &$(32,64,128)$   &   -0.881(75)         \\
\hline \hline
dG$^{(\sigma=1)}$  &$(16,32,64)$    &   0.954(66)        \\
                   &$(24,48,96)$    &   -0.036(23)       \\
                   &$(32,64,128)$   &   -0.309(23)         \\
\hline \hline
dG$^{(\sigma=2)}$  &$(12,24,48)$    &   -0.735(16)          \\
                   &$(16,32,64)$    &   -0.766(16)          \\
                   &$(24,48,96)$    &   -0.882(60)        \\
                   &$(32,64,128)$   &   -0.867(56)          \\
\hline \hline
P   &$(12,24,48)$    &   -1.120(6)          \\
    &$(16,32,64)$    &   -1.089(10)          \\
    &$(24,48,96)$    &   -1.071(42)         \\
    &$(32,64,128)$   &   -0.970(37)         \\
\end{tabular}
\end{ruledtabular}
\end{table}

\section{Raw results from the quotients method}

The effective, lattice-size dependent estimations of critical
points, critical exponents, and dimensionless universal constants
are reported on Table~\ref{tab:FSS_full}. See main text for
definitions and details on our phenomenological renormalization
method.

\begin{table*}
\caption{\label{tab:FSS_full}Crossing points, universal ratios,
and effective critical exponents of the $D=3$ RFIM as given from
the application of the quotients method for all four types of
distributions considered.}
\begin{ruledtabular}
\begin{tabular}{llccccccccc}
Distr. & $(L_{1},L_{2})$ & crossings & $U_{4}$ & $\xi/L$ & $\nu$ & $\xi^{\rm (dis)}/L$ & $\nu^{\rm (dis)}$ & $\eta$ & $\bar{\eta}$ & $2 \eta-\bar{\eta}$ \\
\hline
G  &$(16,32)$   &   2.2779(3)   & 1.0030(1)   &   1.937(3)      &  1.48(3)  & 7.57(2) & 1.36(3)  & 0.5168(6) & 1.0298(2) & 0.0038(11)  \\
   &$(24,48)$   &   2.2758(2)   & 1.0028(1)   &   1.966(3)      &  1.45(3)  & 7.75(2) & 1.39(4)  & 0.5155(5) & 1.0289(2) & 0.0022(11) \\
   &$(32,64)$   &   2.2745(1)   & 1.00261(4)  &   1.989(3)     &  1.36(4)  & 7.88(2) & 1.43(5)  & 0.5150(5) & 1.0280(1) & 0.0019(10) \\
   &$(48,96)$   &   2.2734(1)   & 1.00263(9)  &   2.019(3)     &  1.43(6)  & 8.04(2) & 1.45(6)  & 0.5154(5) & 1.0275(2) & 0.0033(9) \\
   &$(64,128)$  &   2.2731(1)   & 1.0025(1)   &   2.028(3)      &  1.38(9)  & 8.07(2) & 1.35(9)  & 0.5142(5) & 1.0271(3) & 0.0014(10) \\
   &$(96,192)$  &   2.2727(1)   & 1.00243(5)  &   2.042(4)    &  1.38(11) & 8.15(2) & 1.34(11) & 0.5144(5) & 1.0268(1) & 0.0021(11) \\
\hline \hline
dG$^{(\sigma=1)}$&$(16,32)$   &   1.9311(4)  & 1.00030(4)  &   4.340(17)      &  3.04(14)  & 21.21(12) & 1.49(3) & 0.5035(7) & 1.0055(1) & 0.0016(15) \\
                 &$(24,48)$   &   1.9751(2)  & 1.00094(1)  &   2.868(7)       &  2.26(9)   & 12.66(5) & 1.42(3)  & 0.5083(7) & 1.0131(1) & 0.0034(14) \\
                 &$(32,64)$   &   1.9874(1)  & 1.00137(2)  &   2.494(4)       &  1.87(8)   & 10.58(3) & 1.46(4)  & 0.5093(7) & 1.0176(1) & 0.0010(13) \\
                 &$(48,96)$   &   1.9947(1)  & 1.00181(2)  &   2.251(3)       &  1.56(9)   & 9.27(2) & 1.38(5)   & 0.5121(7) & 1.0217(1) & 0.0026(14) \\
                 &$(64,128)$  &   1.9968(1)  & 1.00205(1)  &   2.170(3)       &  1.67(12)  & 8.83(2) & 1.39(5)   & 0.5125(8) & 1.0235(1) & 0.0015(17)  \\
\hline \hline
dG$^{(\sigma=2)}$&$(16,32)$   &   1.0741(5)  & 1.00275(4)  &   1.962(4)      &  1.48(5)   & 7.71(2) & 1.36(4)  & 0.5154(6) & 1.0287(1) & 0.0020(12) \\
                 &$(24,48)$   &   1.0722(4)  & 1.00265(4)  &   1.976(3)      &  1.50(6)   & 7.80(2) & 1.44(6)  & 0.5151(7) & 1.0282(2) & 0.0020(13) \\
                 &$(32,64)$   &   1.0708(3)  & 1.0026(1)   &   1.989(3)      &  1.41(8)   & 7.87(2) & 1.48(9)  & 0.5142(7) & 1.0281(2) & 0.0004(13) \\
                 &$(48,96)$   &   1.0683(3)  & 1.00250(5)  &   2.017(4)      &  1.36(10)  & 8.03(2) & 1.36(9)  & 0.5148(7) & 1.0273(2) & 0.0024(14) \\
                 &$(64,128)$  &   1.0671(2)  & 1.00244(2)  &   2.038(3)      &  1.31(11)  & 8.13(2) & 1.31(11) & 0.5154(6) & 1.0267(1) & 0.0041(13)  \\
\hline \hline
P  &$(16,32)$   &   1.7734(2)   & 1.00363(2) &   1.794(3)      &  1.20(2)    & 6.71(2)  & 1.09(2)  & 0.5183(9) & 1.0373(1) & -0.0006(19) \\
   &$(24,48)$   &   1.7659(2)   & 1.00310(5) &   1.894(4)      &  1.26(3)    & 7.29(2)  & 1.26(3)  & 0.5168(8) & 1.0325(2) & 0.0011(17) \\
   &$(32,64)$   &   1.7625(2)   & 1.00278(2) &   1.955(4)      &  1.30(4)    & 7.64(2)  & 1.34(4)  & 0.5153(8) & 1.0301(1) & 0.0005(17) \\
   &$(48,96)$   &   1.7605(1)   & 1.00259(4) &   2.005(3)      &  1.37(7)    & 7.92(2)  & 1.41(6)  & 0.5143(9) & 1.0282(1) & 0.0004(18) \\
   &$(64,128)$  &   1.7596(1)   & 1.00248(3) &   2.030(3)      &  1.33(7)    & 8.06(2)  & 1.35(7)  & 0.5148(8) & 1.0273(1) & 0.0024(16) \\
   &$(96,192)$  &   1.7589(1)   & 1.00247(9) &   2.060(5)      &  1.43(13)   & 8.22(3)  & 1.48(13) & 0.5146(8) & 1.0266(1) & 0.0026(17) \\
\end{tabular}
\end{ruledtabular}
\end{table*}


\begin{thebibliography}{99}
\bibitem{JOINT} Joint fits share the value of some fitting parameters such as
  the $L\to\infty$ extrapolation (which is the same for all random-field
  distributions), or the corrections-to-scaling exponent
  $\omega$ (which is common to all magnitudes).
\bibitem{CRITICAL} Scaling corrections for the critical point are of order
  $L^{-(\omega+\frac{1}{\nu})}$, $L^{-(2\omega+\frac{1}{\nu})}$,
  etc.~\cite{amit05,prepare}.
\bibitem{yllanes11} The two-exponent scaling relation
  $\eta=(4-D+2\beta/\nu)/2$ yield $\eta=0.5111(8)$,
  see~\cite{fernandez11}. However, this error is underestimated, as it does
  not consider scaling-corrections, nor the uncertainty in the critical value
  of $\hat m$.
\end{thebibliography}
\end{document}